\documentclass[twocolumn,preprintnumbers]{revtex4}

\usepackage{graphicx}
\usepackage{dcolumn}
\usepackage{bm}
\usepackage{color}

\begin{document}

\preprint{}

\title{Photo-induced insulator-metal transition of a spin-electron coupled system}

\author{W. Koshibae$^{1}$}
\author{N. Furukawa$^{2,3}$}
\author{N. Nagaosa$^{1,4,5}$}
\affiliation{%
$^1$Cross-Correlated Materials Research Group (CMRG), RIKEN-ASI, Saitama 351-0198, Japan
\\
$^2$Aoyama-Gakuin University, 5-10-1, Fuchinobe, Sagamihara, Kanagawa 229-8558, Japan
\\
$^3$ERATO-Multiferroics, JST, c/o Department of Applied Physics, 
The University of Tokyo, Tokyo 113-8656, Japan
\\
$^4$Department of Applied Physics, The University of Tokyo, Tokyo 113-8656, Japan
\\
$^5$Correlated Electron Research Group (CERG), RIKEN-ASI, Wako 
351-0198, Japan
}%
\begin{abstract}
The photo-induced metal-insulator transition 
is studied by the numerical simulation of 
real-time quantum dynamics of a double-exchange model.
The spatial and temporal evolutions of the 
system during the transition have been revealed including   
(i) the threshold behavior with respect to the intensity and
energy of light, (ii) multiplication of particle-hole (p-h) pairs 
by a p-h pair of high energy,
and  (iii) the space-time pattern formation such as (a) the stripe 
controlled by the polarization of light, (b) coexistence of  
metallic and insulating domains, and (c) dynamical 
spontaneous symmetry-breaking associated with the
spin spiral formation imposed by the conservation of total spin 
for small energy-dissipation rates. 
\end{abstract}

\maketitle
Changes of electronic states induced by the photo-irradiation 
have been the subject of intensive studies in 
chemistry, biology and physics. 
For example, the photo-induced chemical reactions in molecules
are the basis of many biological functions. 
In solids, the structural changes induced 
by the absorption of photons and the relaxation after it
are a common phenomenon~\cite{Toyozawa}. 
In particular, in correlated electrons such as the transition metal oxides,
the collective nature of the system enhances the sensitivity to 
the external stimuli including photons, and hence even the 
weak photo-irradiation can trigger the changes of 
whole system as the phase transition~\cite{spincross,munekata,koshihara,miyano,fiebig}. 
This is because in strongly correlated electronic systems, 
the rich phases of spin/charge/orbital orderings
with nearly degenerate energies compete with each other~\cite{Imada}. 
For example, in the case of manganites, 
colossal magneto-resistance effect occurs 
because the weak external magnetic field drives a transition
from the insulating charge/spin/orbital ordered state to
the ferromagnetic metallic state~\cite{Tokura}, and  
similar transition can be triggered also by the 
photo-irradiation~\cite{miyano}.

In contrast to these ample examples of 
photo-induced phase transitions, the theoretical studies
are still in the premature stage. Although a classical 
model for the photo-induced phase transition 
has been proposed~\cite{NagaosaOgawa}, the 
quantum theory of photo-induced dynamics of 
correlated electrons are limited to 
small size systems or 
to one-dimensional systems~\cite{matsueda1,matsueda2,yonemitsu}.
Recently, we have proposed a model for the 
real-time quantum dynamics of correlated electrons,
i.e., a double exchange model~\cite{deGennes} where the electrons are interacting
with the classical spins, to simulate the
relaxation after the photo-excitation~\cite{kfn}. 
In that work, we have developed a theoretical method for 
the fully quantum-mechanical time-evolution of 
electronic wavefunction combined with the 
Landau-Lifshitz-Gilbert (LLG) equation for
classical spins, which enables the analysis of the larger-scale/higher-dimensional 
systems.   
By that theoretical study, the self-organized 
space-time structure induced by
the quantum transitions during the relaxation
has been revealed~\cite{kfn}. 

In this paper, we present the real-time quantum dynamics
of the photo-induced insulator-to-metal (IM) transition
for a generalized double exchange model showing the 
competing antiferromagnetic insulating (AFI) and 
ferromagnetic metallic (FM) states separated by the 
first-order phase boundary. 
Many unexplored issues can be addressed in this model, i.e,
the threshold behavior with respect to the 
intensity/energy of light, the effect of the polarization of light,
space-time pattern formation during the transition, 
the role of the conservation of total spins, and 
the asymmetry between the two directions of the 
transition. 

\begin{figure}[b]
\includegraphics[width=70mm,clip]{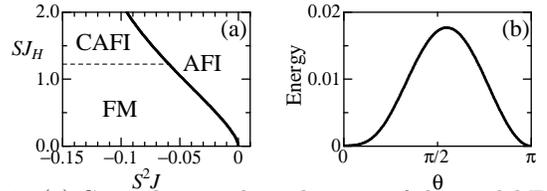}
\vspace{-10pt}
\caption{\label{barrier}
(a) Ground state phase diagram of the model Eq.~(\ref{Hamiltonian}) 
in the plane of $S^2J$ and $SJ_H$ with 
fixed $S^4J_N$=$-0.2$ in units of $t$.  There exist three phases, i.e., 
ferromagnetic metal (FM), antiferromagnetic insulator (AFI) and 
canted antiferromagnetic insulator (CAFI).  
(b) Lowest energy per site measured from ground state 
as a function of angle between neighboring local-spins 
in a two sub-lattice consideration on $8\times 8$ system (see text).  
}
\end{figure}
 
We start with the Hamiltonian, 
\begin{eqnarray} 
\label{Hamiltonian}
\hat H&=&-t\sum_{<ij>,s}c^\dagger_{is}c_{js}+h.c.
-J_H\sum_{iss'}
c^\dagger_{is}c_{is'}\vec{\sigma}_{ss'}
\cdot \vec{S}_i \nonumber
\\ 
&&+J\sum_{<ij>}\vec{S}_i\cdot \vec{S}_j
+J_N\sum_{<ij>}\left(\vec{S}_i\cdot \vec{S}_j\right)^2
\end{eqnarray} 
where $<$$ij$$>$ denotes a nearest-neighbor pair, $s$ and $s'$ 
are indices for electron spin, respectively,   
and $\vec{\sigma}=(\sigma^x,\sigma^y,\sigma^z)$ are Pauli matrices.  
The local spins, $\vec{S}_i$'s, are taken to be classical vectors with 
magnitude $S$, and other notations are standard.  
We consider the half-filled case, i.e., one electron per site.  
In Eq.(\ref{Hamiltonian}), we have introduced the biquadratic exchange 
interaction with coupling constant $J_N$, 
which gives an energy barrier between 
ferromagnetic (F) and antiferromagnetic (AF) states
when $J_N$ is negative. 

Figure 1(a) shows the ground state phase diagram of the model Eq.~(\ref{Hamiltonian}) 
on the two-dimensional square lattice, 
in the plane of $S^2J$ and $SJ_H$ with fixed $S^4J_N$=$-0.2$ in units of $t$.  
The solid curve indicates the boundary of first-order 
phase transition. In the right side, the antiferromagnetic
spins open a gap and the system is insulating. In the left side,
the spins are perfectly or almost ferromagnetic.  In the ferromagnetic state, 
the system is metallic, but 
the small staggered component opens a small gap for $SJ_H\!\!\!\!$
\hspace{0.3em}\raisebox{0.4ex}{$>$}\hspace{-0.75em}\raisebox{-.7ex}{$\sim$}\hspace{0.3em}
$\!\!\!\!1.2$~\cite{foot}. 
The phase transition at  
$SJ_H$$\simeq$1.2 (broken line) is of the continuous second order.  

Let us focus on the boundary of first-order 
phase transition.  
Figure \ref{barrier}(b) shows the energy 
as a function of the angle $\theta$ between the sublattice
spin directions for $8\times8$ size system in periodic boundary condition 
with a parameter set, 
$t$=1, $SJ_H$=1, $S^2J$=$-0.043$, $S^4J_N$=$-0.017$,  
$S$=1.  This condition is used hereafter.  
The AF state and the F state, 
respectively given by $\theta=\pi$ and 0,
are almost degenerate, and show the local stabilities with the
potential barrier between them.
  
We introduce the 
Landau-Lifschitz-Gilbert (LLG) equation for the motion of the local spins, 
$
\dot{\vec{S}}_i=
\vec{h}_{eff,i}\times\vec{S}_i
-\alpha\vec{S}_i\times\dot{\vec{S}}_i,
$
where $\vec{h}_{eff,i}=\left< \Phi(T) | -\partial \hat{H}/\partial{\vec{S}}_i
| \Phi(T) \right>$ with $\left.|\Phi(T)\right>$ being the electronic wavefunction 
of a Slater determinant at time $T$ obtained by solving the
Schr\"{o}dinger equation as discussed in Ref.~\cite{kfn}.
The Gilbert damping constant $S\alpha$ 
describes all the other relaxation processes 
of the spins than the coupling to the conduction electrons.  

\begin{figure}[t]
\includegraphics[height=85mm,angle=270,clip]{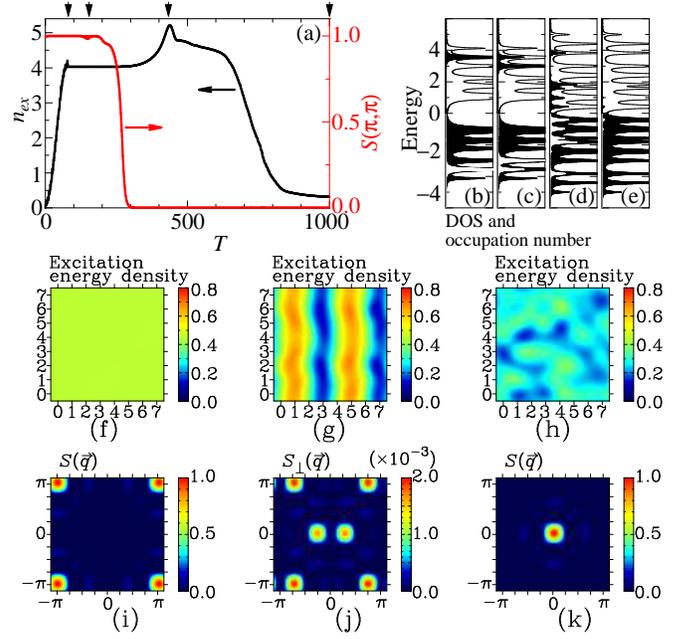}
\begin{tabular}{ccc}
\begin{minipage}{0.33\hsize}
\includegraphics[height=26mm,angle=270,clip]{80EMf.eps}
\end{minipage}
\begin{minipage}{0.33\hsize}
\includegraphics[height=26mm,angle=270,clip]{152EMf.eps}
\end{minipage}
\begin{minipage}{0.33\hsize}
\includegraphics[height=26mm,angle=270,clip]{430EMf.eps}
\end{minipage}
\end{tabular}
\begin{tabular}{ccc}
\begin{minipage}{0.33\hsize}
\includegraphics[height=26mm,angle=270,clip]{80EMs.eps}
\end{minipage}
\begin{minipage}{0.33\hsize}
\includegraphics[height=26mm,angle=270,clip]{152EMs.eps}
\end{minipage}
\begin{minipage}{0.33\hsize}
\includegraphics[height=26mm,angle=270,clip]{430EMs.eps}
\end{minipage}
\end{tabular}
\vspace{-5pt}
\caption{\label{largea}
(Color online) Time evolution of the photo-excited electronic state 
and its relaxation dynamics 
for $S\alpha=1$.  
(a) Black line with scale on left axis 
is the time dependence of total occupation number above Fermi level, $n_{ex}$.  
Red (gray in print) line with scale on right axis 
shows the time dependence of the equal-time spin-structure-factor, 
defined by 
$S(\vec{q})
={1\over N}\sum_{ij}\vec{S}_i\cdot\vec{S}_je^{i\vec{q}\cdot\left(\vec{i}-\vec{j}\right)}$, 
for $\vec{q}=(\pi,\pi)$, where $\vec{i}$ and $\vec{j}$ 
are the vector representation 
of the lattice points $i$ and $j$, respectively, 
$\vec{q}$ is that of the reciprocal lattice-point, 
and $N$ denotes the system size.     
The arrows on top axis are the marks for $T=80, 152, 430$ and 1000, respectively.  
(b) Electron occupancy at $T=80$ represented by the black area 
with the density of states (DOS) (thin line).  
The Fermi energy is taken to be zero.  
In the same way, the electron occupancy at 
$T=152$, 430 and 1000 are shown in (c), (d) and (e), respectively.    
Figures (f), (g) and (h) are 
the distribution of excitation-energy-density in real space 
measured from the ground state at $T=80$, 152, and 430, 
respectively.  
Numbers on horizontal (vertical) axis is for  
the coordinates of lattice points in $x$ ($y$) direction.    
(i) $S(\vec{q})$ at $T=80$. 
(j) Transverse component of the equal-time spin-structure-factor $S_\bot(\vec{q})$ 
at $T=$ 152.  
(k) The same with (i) but $T=430$.  
A spline interpolation is used 
for figures (f)-(k).  
}
\end{figure}

Figure \ref{largea} shows time-evolution 
of the photo-excited electronic state and its relaxation dynamics 
for $S\alpha=1$.  The unit of time $T$ is $\hbar/t$ and hence is
typically $\sim 10^{-15}$ sec assuming $t = 0.4$eV.   
We start with the AFI state, i.e., 
the electronic state has an insulating energy-gap $2SJ_H$  
and lower energy band is occupied by electrons and 
the upper energy band is empty at $T=0$.  
In order to mimic the thermal fluctuation, we introduce a 
random tilting of each spin from the perfect AF configuration 
up to 0.1 rad which corresponds to the state with an
excitation energy of $\sim0.0003t$ from the ground state.  
The photo-excitation is achieved by introducing the time-dependent 
vector potential into the hopping matrix element:  
for the nearest-neighbor pair $<$$ij$$>$ in $y$-direction, we use    
$t\rightarrow te^{iA(T)}$ where the phase $A(T)$ of the hopping matrix element  
is given by the vector potential.   
In the early stage, $T=0\sim80$, we have applied  
$A(T)=0.1\sin(\omega T)$ in this simulation.  
The frequency $\omega$ is tuned for the energy difference between 
the second highest and the second lowest energy levels of the AF state 
on the finite size system.  
The particles are excited from lower to upper energy band (see Fig.~\ref{largea}(a)) 
and the particle-hole (p-h) pairs are created on 
the second highest and the second lowest energy levels (see Fig.~\ref{largea}(b)).  
In the excited state, the spatial distribution of excitation energy is uniform 
as seen in Fig.\ref{largea}(f).  
At $T=80$, total amount of the excited electron is about 4, and after that, $A(T)=0$.  

Up to $T\sim200$, 
the local-spin structure is almost AF one, so that 
the density of states (DOS) hardly changes  
as shown by thin lines in Fig.~\ref{largea}(b) and (c).  
However, the electronic state has been changed dramatically, i.e., 
the electron occupation shows a time evolution and 
the spatial distribution of excitation energy forms 
a {\it stripe} pattern as seen in Figs.~\ref{largea}(c) and (g).  
Correspondingly, 
the transverse component of the local-spins 
shows a dynamical pattern as in Fig.~\ref{largea}(j).  
During the photo-excitation by the light polarized in $y$ direction, 
the electrons are accelerated along 
the direction.  The polarized motion of electrons derives  
a deformation of the local-spin structure to be favorable for the electron motion.    
The characteristics of the relaxation dynamics appear in the {\it stripe} pattern 
as seen in Fig.~\ref{largea}(g) reflecting the wave function and the 
spatial distribution of the excited p-h pairs.

In the period $T=200\sim300$, the system shows a drastic change, that is, 
AFI to FM transition.  
For 
$T<200$, the spin-structure-factor 
$S(\pi,\pi)$ is almost 1, and 
we clearly see the insulating energy-gap $2SJ_H$ between upper 
and lower energy bands, 
i.e., the system is the (excited) AFI.  
For  
$T>300$, on the other hand, $S(\pi,\pi)$ is almost 0, 
and the gap closes.  
In fact, the energy level structure is very close to 
that of the tight-binding Hamiltonian for free-electrons with a Zeeman splitting.  
(The sparse energy level distribution around zero in energy axis is  
due to the finite-size effect.)  
As seen in Fig.~\ref{largea}(k), the local-spin becomes the
F state corresponding to the metallic electronic-states.  
The electrons, however, remain still highly excited.  
As seen in Figs.~\ref{largea}(a) and (d), the total number of excited 
electrons above Fermi level increases and shows a peak at $T\sim430$. 
After that, it decreases but 
is as large as the number of initially excited electrons, 
up to $T\sim660$.  
On the other hand, it has been confirmed that the total energy of the system, i.e., 
the electrons and the local spins, 
decreases monotonically due to the Gilbert damping.

The enhancement of the total number of excited electrons above Fermi level 
is due to the {\it Auger process}.  
This is a characteristics of the interacting electron systems, 
and is in sharp contrast to the relaxation dynamics of conventional 
semiconductors which physics is well described by the single
carrier problem.   
This multiplication of p-h pairs during the relaxation dynamics 
offers an interesting 
possibility to enhance the photo-current generation and solar cell reaction.

\begin{figure}[t]
\includegraphics[height=80mm,angle=270,clip]{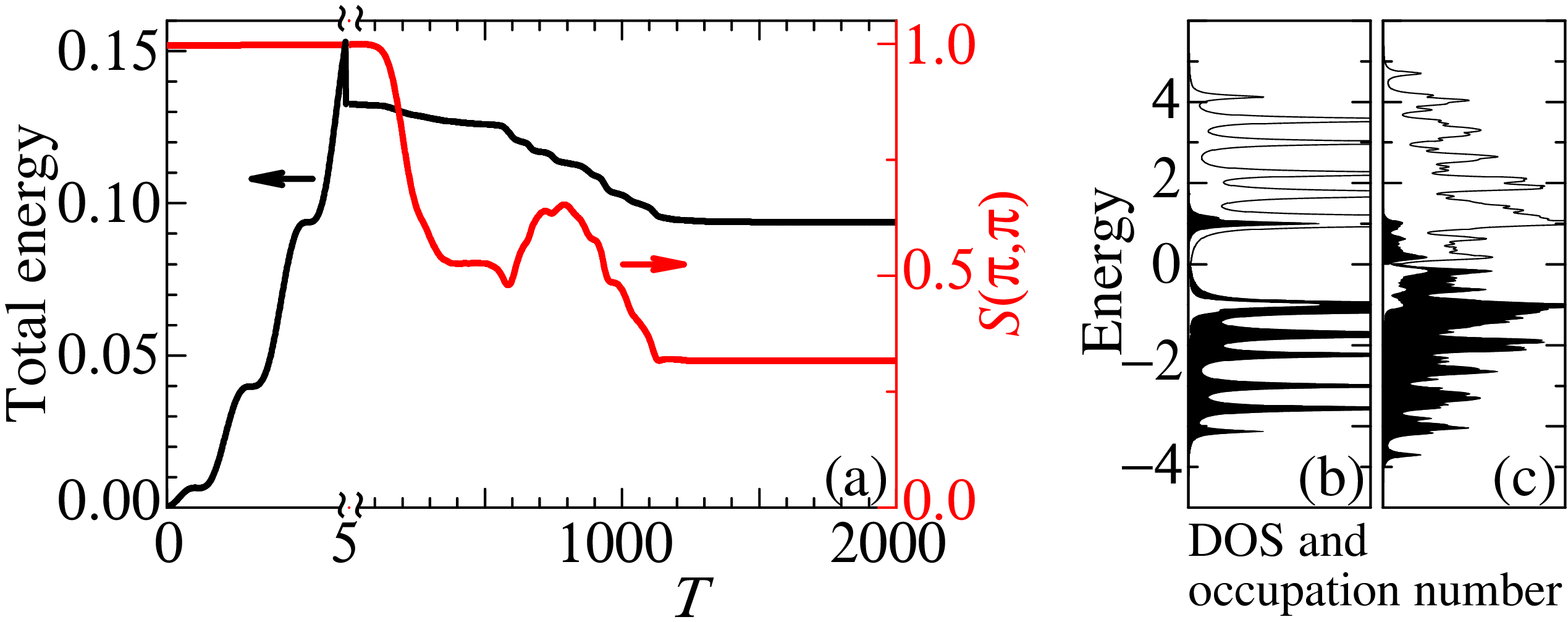}
\begin{tabular}{cc}
\begin{minipage}{0.67\hsize}
\includegraphics[width=60mm,clip]{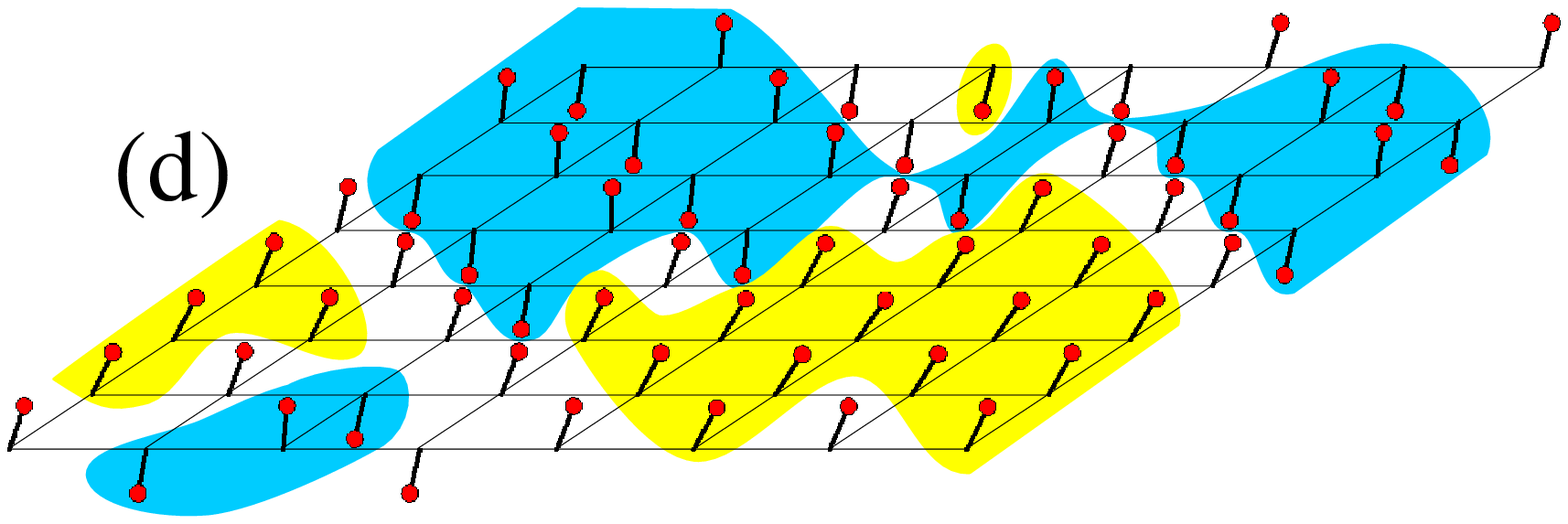}
\end{minipage}
\begin{minipage}{0.33\hsize}
\includegraphics[height=26mm,angle=270,clip]{2000EMf.eps}
\end{minipage}
\end{tabular}
\vspace{-5pt}
\caption{\label{domain}
(Color online) 
(a) Black line with scale on left axis 
is the time dependence of total energy per site measured from ground state.    
Red (gray in print) line with scale on right axis 
shows $S(\pi,\pi)$ as a function of $T$.   
For the photo-excitation, $A(T)=0.15\sin(2SJ_HT)$ is used.  
Different time scales are used for $T<5$ and $5<T$, 
respectively, to zoom up the initial stage for the photo-excitation. 
During the photo-excitation, total energy increases, and later on ($5<T$) it decreases 
due to the Gilbert damping.  
(b) The same with Fig.~\ref{largea}(b) but $T=6$ in this simulation.    
(c) The same with (b) but $T=2000$.    
(d) Local spin structure at $T=2000$.  The stick with dot indicates the local spins.  
There exist AF and F domains.  
The F and AF domains are expressed by yellow (gray) and blue (dark gray) regions, respectively.  
In the figure, the spin belongs to F (AF) domain 
when all the neighboring spins are in F (AF) alignment, 
and the other spins are put into the white area.    
(e) The same with Fig.~\ref{largea}(f) but $T=2000$ in this simulation.
}
\end{figure}

This AFI-to-FM transition, however, does not occur when the excitation is not strong enough.  
For the results shown in Fig.~\ref{largea}, 
the photo-excitation has been done in the early stage,  
$0\leq T\leq T_f=80$, with $\omega \cong 7SJ_H$.  
By decreasing the period $T_f$, we can reduce the number of 
the p-h pairs by the photo-excitation $n_{ex}(T_f)$.  
We find that the IM transition hardly occurs for $n_{ex}(T_f)$
\hspace{0.3em}\raisebox{0.4ex}{$<$}\hspace{-0.75em}\raisebox{-.7ex}{$\sim$}\hspace{0.3em}
2, in the present numerical condition.  

We have also examined the frequency $\omega$ dependence for the IM transition.  
In the case that $\omega=2SJ_H$, the threshold of the IM transition is lying on 
$4<n_{ex}(T_f)<5$ in the numerical simulations.  
Although the IM transition occurs even in this case, 
we do not find the Auger process.  
Comparison with the above case of $\omega \cong 7SJ_H$, 
it is concluded that the higher energy photon is more effective to induce 
the IM transition through the Auger process.

When $n_{ex}(T_f)$ is close to the threshold,   
we find the coexistence of the insulating and metallic 
domains.  
The results for $\omega=2SJ_H$ and $n_{ex}(T_f)\simeq4.13$ 
are summarized in Fig.~\ref{domain}.   
The local-spin state changes gradually in a long time period, 
150
\hspace{0.3em}\raisebox{0.4ex}{$<$}\hspace{-0.75em}\raisebox{-.7ex}{$\sim$}\hspace{0.3em}
$T$
\hspace{0.3em}\raisebox{0.4ex}{$<$}\hspace{-0.75em}\raisebox{-.7ex}{$\sim$}\hspace{0.3em}
1100 (see the time dependence of $S(\pi,\pi)$ in Fig.~\ref{domain}(a)).   
A long time later, 
the AF (blue) and F (yellow) domains are spatially 
separated as seen in Fig.~\ref{domain} (d),
and the corresponding excitation energy density is shown in 
Figs.~\ref{domain} (e).
The energy levels and occupation numbers before and after the
transition are shown in Figs.~\ref{domain}(b) and (c), respectively.
The residual p-h pairs in Fig.~\ref{domain} (c)
are confined in the F metallic domain, 
so that the electronic state 
form a slightly higher energy density region 
on the F domain and the stable AFI domain 
(see Figs.~\ref{domain}(d) and (e)).     
This meta-stable state continues for 
a long time within our simulation (at least up to $T\sim6000$).

\begin{figure}[t]
\includegraphics[height=85mm,angle=270,clip]{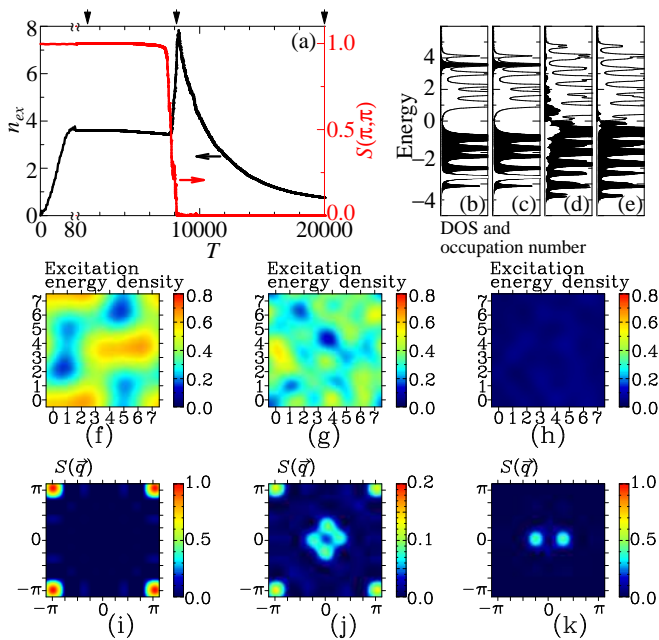}
\begin{tabular}{ccc}
\begin{minipage}{0.33\hsize}
\includegraphics[height=26mm,angle=270,clip]{1000EMf.eps}
\end{minipage}
\begin{minipage}{0.33\hsize}
\includegraphics[height=26mm,angle=270,clip]{8110EMf.eps}
\end{minipage}
\begin{minipage}{0.33\hsize}
\includegraphics[height=26mm,angle=270,clip]{20000EMf.eps}
\end{minipage}
\end{tabular}
\begin{tabular}{ccc}
\begin{minipage}{0.33\hsize}
\includegraphics[height=26mm,angle=270,clip]{1000EMs.eps}
\end{minipage}
\begin{minipage}{0.33\hsize}
\includegraphics[height=26mm,angle=270,clip]{8110EMs.eps}
\end{minipage}
\begin{minipage}{0.33\hsize}
\includegraphics[height=26mm,angle=270,clip]{20000EMs.eps}
\end{minipage}
\end{tabular}
\vspace{-5pt}
\caption{\label{smalla}
(Color online)   
The same with Fig.~\ref{largea} but $S\alpha=0.01$ is used. 
In (a), different time scales are used for $T<80$ and $80<T$, 
respectively, to zoom up the initial stage.    
The arrows on top axis are the marks for $T=1000, 8110$ and 20000, respectively.  
Figure (b) is the same with Fig.~\ref{largea}(b) but $S\alpha=0.01$.  
The figures (c), (f) and (i) show the results at $T$=1000.  
The figures (d), (g) and (j) show the results at $T$=8110.  
Here, $S(\vec{q})$ is presented on (j), instead of $S_\bot(\vec{q})$.   
The figures (e), (h) and (k) show the results at $T$=20000.  
}
\end{figure}

We have also studied the more realistic case of $S\alpha$=0.01 (Fig.~\ref{smalla}).  
Other parameters are the same with the case shown in Fig.~\ref{largea}.  
Due to the reduced relaxation rate, it takes  
much longer time for whole relaxation dynamics compared with that in Fig.~\ref{largea}.
Despite the reduced relaxation rate, the real-time dynamics of 
the electronic and local-spin states is not silent and is rather active.   
During the relaxation dynamics, the direction of the staggered 
magnetization changes drastically.  
And the spatial inhomogeneity of the electronic state has also been developed 
(see Fig.~\ref{smalla}(f)).  
The slow energy relaxation is advantageous for the 
Auger process resulting in the large amount of p-h pair creation 
at $T\sim8000$ when the gap closing occurs (see Fig.~\ref{smalla}(a)).  
At this transition, all the local-spins play very active dynamics with 
developing spatial inhomogeneity of the electronic state  
 (see Fig.~\ref{smalla}(g)).  
The change of spin configuration from AF state 
is clearly seen in the equal-time spin-structure-factor 
shown in Figs.~\ref{smalla}(a) and (j): 
The intensity at $(\pi,\pi)$ becomes small 
and that at around the center region is grown. 
Near the center, there exist four spots, and two of them 
disappear later on in the time evolution.   
Finally, a {\it spiral} spin-structure is obtained in this case 
(see Fig.~\ref{smalla}(k)).   
Due to the Gilbert damping $S\alpha$, 
the total spin-angular-momentum can change  
in the relaxation dynamics.  
For large $S\alpha$ the spin state can be easily changed from AF to 
F as in Fig.~\ref{largea}(g) but not for small $S\alpha$ 
because the conservation law becomes effective. 
Therefore, the ferromagnetic state is avoided and 
the {\it spiral} spin configuration appears characterized by  
the equal-time spin-structure-factor shown in Fig.~\ref{smalla}(k).  
In the present case, the propagation vector of the {\it spiral} spin-state 
is in $x$ direction. This is the dynamical symmetry breaking, and 
the direction ($x$ in this particular case) is chosen by the 
accidental reason while the memory of the polarization of photo-excitation ($y$ direction) 
has been already lost.

So far, we have discussed the AFI-to-FM transition.  
As shown in Fig.~\ref{barrier}(b), FM and AFI states are almost degenerate, 
so that FM-to-AFI transition of this system 
is worth to be examined.  
In the F state, however, the excitation with $\vec{q}=0$ is prohibited.  
Therefore, we have prepared the initial excited-state within F state by 
putting the initial electron occupation by "hand".  
Although we have examined a number of numerical simulations, 
metal-to-insulator transition has not been obtained.  
The numerical results suggest that 
metal-to-insulator transition 
is much more difficult 
than insulator-to-metal transition in the present model.   

In summary, we have studied the insulator-to-metal 
transition by the photo-irradiation in a model 
of competing AF and F states. 
(i) The threshold behavior with respect to the intensity and
energy of lights, (ii) Auger process, i.e., multiplication 
of particle-hole (p-h) pairs by a p-h pair of high energy,  
(iii) the space-time pattern formation including the stripe
controlled by the polarization of light, 
and (iv) the intriguing spin spiral state by the 
conservation of total spin for weak Gilbert damping 
have been revealed. The photo-carrier generation under 
the external electric field and its implications
to the solar cell reaction are the issues left for future
studies.

The authors are grateful to Y. Tokura, M. Kawasaki, H. Matsueda, T. Tohyama, 
S. Ishihara, and K. Tsutsui for useful discussions.
This work is supported by Grant-in-Aid for Scientific Research 
(Grant No. 19048015, 19048008, 17105002, 21244053, and 21360043)
and a High-Tech Research Center project for private universities 
from the Ministry of Education, Culture, Sports, Science and Technology of
Japan, Next Generation Supercomputing
Project of Nanoscience Program, JST-CREST and NEDO.  
N.N. is supported also by Strategic International Cooperative 
Program (Joint Research Type) from JST, and 
Funding Program for World-Leading Innovative R $\&$ D on 
Science and Technology (FIRST Program).


\begin{thebibliography}{}
\bibitem{foot}
For $SJ_H > 4t$, the system is insulating since the energy bands  
for up and down spin split into two bands separated by a 
gap $2(SJ_H - 4t)$ in F state.
\end{thebibliography}
\end{document}